\documentclass[final, 2p, times, twocolumn]{elsarticle}
\usepackage{amssymb}
\usepackage{amsmath}
\biboptions{sort&compress}

\usepackage[version=4]{mhchem}
\usepackage{booktabs} 
\usepackage{enumitem}
\usepackage{refcount}
\newcommand{\cmmnt}[1]{}

\usepackage{graphicx}
\usepackage{ctable} 
\usepackage{array, boldline, makecell, booktabs}
\usepackage{flexisym}
\usepackage{float} 
\usepackage{caption}
\usepackage{subcaption}

\usepackage{amsmath,amsfonts,mathtools}

\journal{Journal of Quantitative Spectroscopy and Radiative Transfer}

\begin{document}

\begin{frontmatter}

\title{\ce{H2}-Induced Pressure Broadening and Pressure Shift in the $P$-Branch of the $\nu_3$ Band of \ce{CH4} from 300 to 700 K}

\author{Ehsan Gharib-Nezhad$^{1}$, Alan N. Heays$^{2, 3}$, Hans A. Bechtel$^{4}$, James R. Lyons$^{2}$}
\address{$^{1}$School of Molecular Sciences, Arizona State University, Tempe, AZ. 85287, USA.}
\address{$^{2}$School of Earth and Space Exploration, Arizona State University, Tempe, AZ. 85287, USA.}
\address{$^{3}$NASA Astrobiology Institute, NASA Ames Research Center, Moffett Field, CA., USA.}
\address{$^{4}$Advanced Light Source, Lawrence Berkeley National Laboratory, Berkeley, CA. 94720, USA.}

\begin{abstract} 
   For accurate modelling of observations of exoplanet atmospheres, quantification of the pressure broadening of infrared absorption lines for and by a variety of gases at elevated temperatures is needed. High-resolution high-temperature \ce{H2}-pressure-broadened spectra are recorded for the \ce{CH4} $\nu_3$-band $P$-branch. Measured linewidths for 116 transitions between 2840 and 3000 cm$^{-1}$ with temperature and pressures ranging between 300 and 700 K, and 10 and 933 Torr, respectively, were used to find rotation- and tetrahedral-symmetry-dependent coefficients for pressure and temperature broadening and pressure-induced lineshifts.   The new pressure-broadening data will be useful in radiative-transfer models for retrieving the properties of observed expolanet atmospheres.
  
\end{abstract}


\end{frontmatter}

\section{Introduction} \label{sec: intro}
Methane (\ce{CH4}) has been observed in the infrared spectra of  different solar-system atmospheres including those of terrestrial planets (e.g., on the surface of Mars \cite{Krasnopolsky2004, Lyons2005}), Jovian planets (e.g., Jupiter, Saturn, Uranus \cite{Orton2014, Orton2014a}), and Titan \cite{Atreya2003, Lodders2010}. The abundance of \ce{CH4} is also important in constraint understanding the C/O ratio in the atmospheres of brown dwarfs and exoplanets, as well as understanding their formation history \cite{Marley2015, Fortney2012_C/Oratios}.  Because the thermochemically dominant carbon-bearing molecule at T$>$1200 K is CO and at T$<$800 K is \ce{CH4} \cite{Lodders2010}, their mixing ratios  with \ce{CO2} are used as a temperature probe and to determine super-Earths/sub-Neptune metallicities \cite{Venot2014,Kreidberg2018}. Moreover, \ce{CH4} near-infrared (NIR) spectra are an important tool for classifying  brown dwarf types (e.g., T-dwarfs \cite{Burgasser2006}).
Despite extensive endeavors to model the chemical composition of exoplanetary atmospheres by means of radiative transfer modeling (i.e., transmission and emission exoplanetary spectra \cite{Line2011Gj436bCH4, Kreidberg2018Wasp107,Swain2010HD189733bCH4}), a proposed detection of \ce{CH4} is still under debate \cite{Desert2009HD189733bCO}. Additionally, high-resolution Earth-based searches of methane through the cross-correlation technique have been unsuccessful \cite{Wang2018_CH4_CCT}. However, it has been argued thermochemically that \ce{CH4} is one of the main absorbers in super-Earth to sub-Neptune atmospheres \cite{Moses2013}. { \ce{H2} is the major broadening molecule (or perturber) in these exoplanetary atmospheres, and therefore,} the accuracy of radiative transfer modeling, particularly for the cross-correlation technique (see section 3.5 in \cite{Brogi2019}),  relies strongly upon the accuracy and completeness of \ce{CH4} spectroscopic data including rovibrational transitions and pressure-broadening coefficients appropriate for high-temperature and { \ce{H2}-dominated} atmospheres \cite{Freedman2014, GharibNezhad2019}. Accurate quantification of pressure-broadening coefficients at room- and high-temperature is fundamental because they influence the absorption cross-section data and, therefore, the modeled exoplanet atmospheric spectra \cite{GharibNezhad2019, Fortney2019whitepaper}.

Methane is a tetrahedral molecule with five symmetry species: $A_1$,  $A_2$, $E$, $F_1$, and $F_2$. The $\nu_3$ fundamental band arises from asymmetric C--H stretching (see chapter 7 in \cite{Bernath_book}). Given the relevance of \ce{CH4} infrared (IR)  absorption and emission in the study of brown dwarfs and planetary/exoplanetary atmospheres, many experimental and theoretical studies recorded or computed the relevant rovibrational transitions. High-resolution IR spectra of \ce{CH4} have been recorded at both room \cite{Albert2009,Campargue2013} and high temperatures \cite{Wong2019, Hargreaves2015, Nassar2003, Perrin2007, Hargreaves2012, Thievin2008}. Additionally, several ab-initio studies have computed the \ce{CH4} rovibrational transitions \cite{Wong2019, Ba2013,Nikitin2015,Rey2017,Rey2018}.

Since the 1980s, several laboratory measurements of the pressure-broadening of \ce{CH4} by various broadeners (hereafter referred to as absorber@[broadener], e.g.,  \ce{CH4}@[\ce{H2}]) at room temperature have been made. Non-Voigt pressure-broadening coefficients of the \ce{CH4}@[\ce{H2}, \ce{N2}, \ce{Ar}, or \ce{He}] $\nu_3$ band $Q$ branch were analyzed using a laser spectrometer at high resolution \cite{Pine1992, Pine2003} and showed a strong dependency of linewidths on broadener and total angular-momentum quantum number, $J$. In addition, the measured linewidths are dependent on the tetrahedral symmetry species (i.e., $A_1, A_2, E, F_1, F_2$). The $R$-branch of the 3$\nu_3$  overtone ($\sim\!9000$ cm$^{-1}$) of \ce{CH4}@[\ce{H2}] was measured up to $J_{\rm{lower}}=J^{\prime\prime}$=6 by Fourier transform spectroscopy with 0.01 cm$^{-1}$ spectral resolution \cite{Fox1985, Fox1988}.

{ Several studies have used quantum or semi-classical approaches to calculate, predict, and explain pressure broadening of \ce{CH4} in different broadeners (or perturbers)\cite{Hartmann2018}. Anderson theory, for instance, utilizes a perturbation approach to compute the line broadening and their temperature-dependence coefficients through electrostatic interactions \cite{Anderson1949,Frost1976,Tejwani1971}. However, it was shown later that electrostatic forces are not able to explain the broadening for some perturbers such as \ce{O2} and \ce{N2} \cite{Devi1983}. In comparison, Robert-Bonamy theory \cite{Gabard2013} was used to show the atom-atom potential energy is the main cause of collisional broadening for these species \cite{Neshyba1994}.} 

To the best of our knowledge, there are no measurements of \ce{CH4}@[\ce{H2}] (or any other broadeners) at T$>$~315 K. Measurements for temperatures between 200 and 300 K show the temperature-dependence coefficient ($n_{\text{$\gamma$}}$, see section~\ref{sec:n-gamma-4-2}) of the $\nu_3$ band of \ce{CH4}@[Air] and \ce{CH4}@[\ce{N2}] is 0.6 -- 1.0 \cite{Varanasi1975} and 0.94--0.97 \cite{Devi1983}, respectively. For the $\nu_4$ band of \ce{CH4}@[Air] and \ce{CH4}@[\ce{N2}], n$_{\text{$\gamma$}}$ is 0.5 -- 0.8 \cite{Smith1992}. A complete list of literature regarding measurement of temperature-dependence coefficients is reported in Table \ref{tab:Literature-nT}. 

For this study, we used Fourier-transform infrared spectroscopy (FTIR) to record rovibrational lines of \ce{CH4}@[\ce{H2}] in the $P$ branch of the $\nu_3$ band over the temperature range 300--700 K (Sec. \ref{sec: exp details}). Then, using a least-squares fitting analysis, the Lorentzian linewidth ($\gamma_{\text{L}}$) and temperature-dependence coefficients ($n_{\text{$\gamma$}}$) are determined for $J^{\prime\prime}$= 2 -- 17 (Sec. \ref{sec: analysis}). The dependency of the Lorentzian coefficients on total quantum number  $J^{\prime\prime}$ and the tetrahedral symmetry species is discussed in Sec.~\ref{sec: results}.

\section{Experimental details} \label{sec: exp details}
\subsection{Instrumental setup}
\begin{figure}[h]
\centering
 \includegraphics[scale=0.5] {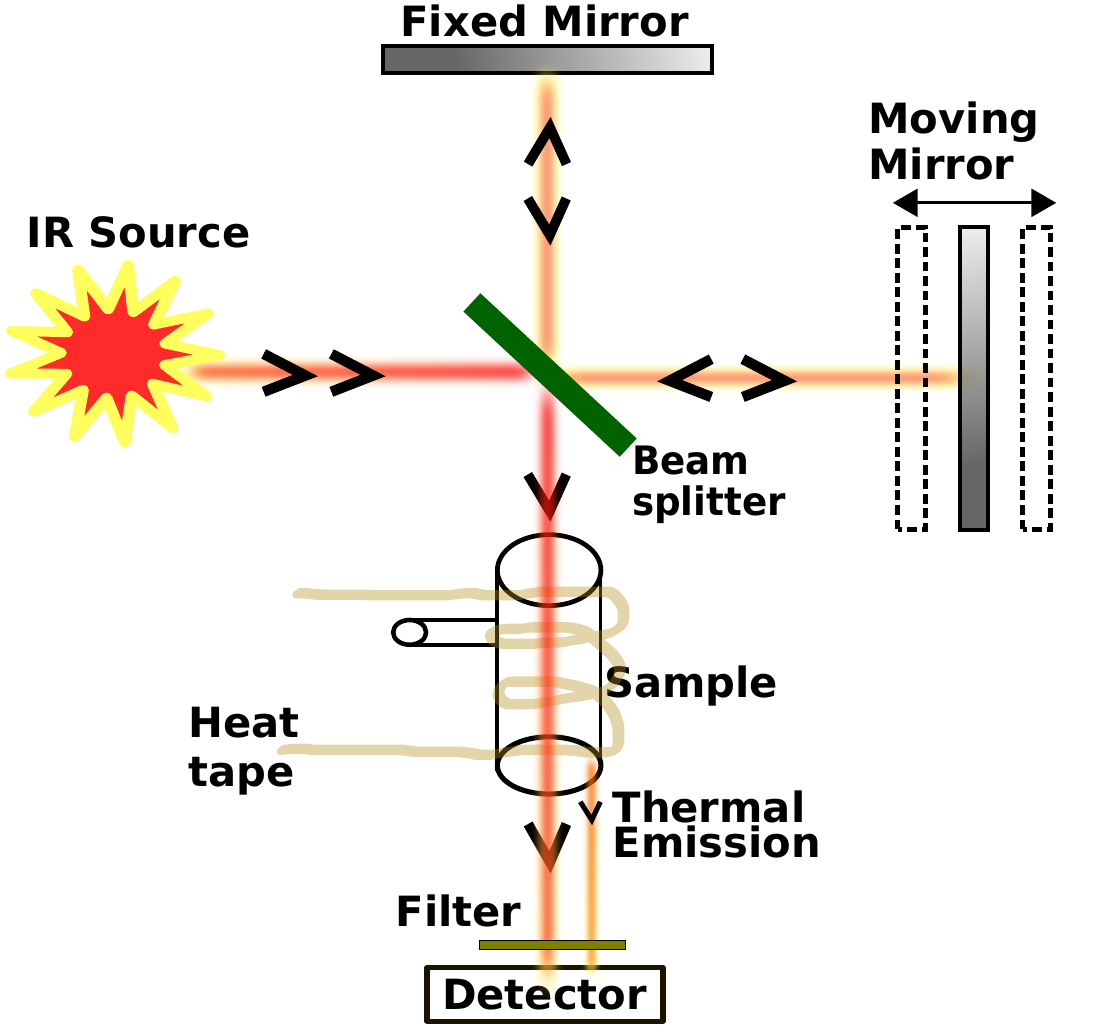} 
	  \caption{Schematic view of the Bruker 125HR IFS spectrometer: both the sample gas cell and the heat tape are located inside the evacuated sample compartment. The infrared emission from the gas cell does not contribute to the interferogram, and the AC-copuled detector also automatically eliminates infrared emission of the gas cell. Additionally, fused silica and Ge filters were used between sample and detector to prevent detector saturation. }
	  \label{fig:setup scetch} 
\end{figure}

All spectra in this study were recorded with a Bruker 125HR infrared Fourier-transform spectrometer located at the Advanced Light Source (ALS) of the Lawrence Berkeley National Laboratory (LBNL). As shown in  Fig.~\ref{fig:setup scetch}, the evacuated sample chamber in this model of spectrometer is located between the beam splitter and detector. In this case, the thermal IR emission from the heated sample gas cells does not contribute to the recorded interferogram, and no post-analysis correction for the cell emission is required in comparison with other studies in which the heated cell was placed at the entrance to the spectrometer (e.g., Ref. \citep{Hargreaves2015}). For measurements at high temperature, we designed a sealed monolithic gas cell. Due to its high transmittance over a spectral range of 2750--3250 cm$^{-1}$ and high melting point, the whole gas cell and spectral windows are fabricated from fused quartz. 

\begin{figure*}[h]
\centering
 \includegraphics[scale=0.98] {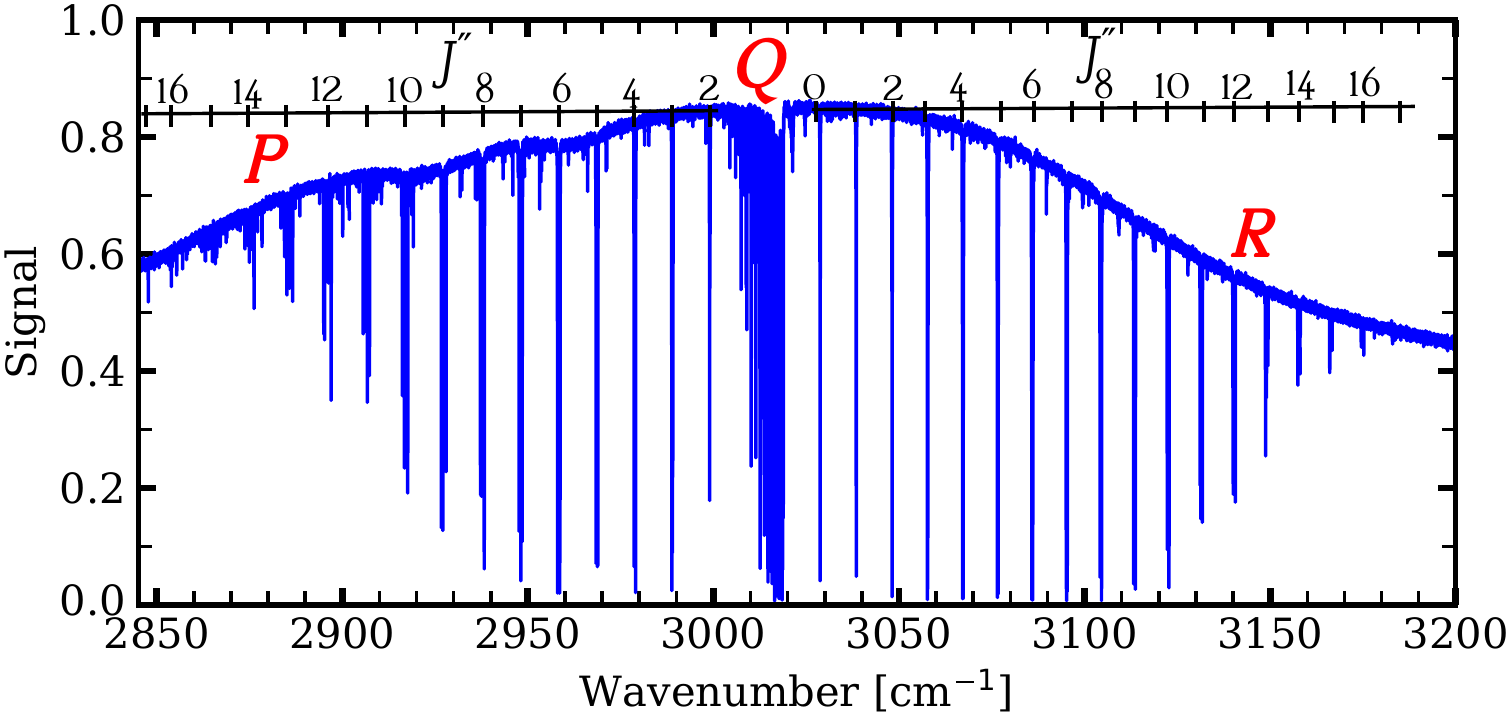} 
	  \caption{The measured spectrum of the \ce{CH4} $\nu_3$ band: $P_{\rm{\ce{CH4}}}$=1.1 Torr and $P_{\rm{\ce{H2}}}$=100.0 Torr. The strong lines ($J^{\prime\prime}$=0--17) belong to the $\nu_3$ band consisting of $P$, $Q$, and $R$ branches. Weak lines belong to $\nu_3+\nu_4-\nu_4$ and $\nu_2+\nu_3+\nu_2$ combination bands. }
	  \label{fig:Spectra-measured} 
\end{figure*}

\subsection{Recorded Spectra}
We recorded spectra for temperatures 300, 500 and 700~K and over a pressure range of $\sim$0.8 -- 7.0 Torr for \ce{CH4} gas and $\sim$10 -- 933 Torr (0.013--1.2 atm) of \ce{H2} broadening gas.  In total, four quartz gas cells with a path length of 10$\pm$0.2 cm were used in these measurements. Different amounts of \ce{CH4} and \ce{H2} gases were inserted in each tube at room temperature and then the port was sealed. After sealing the gas cell at room temperature with known number densities of {\ce{CH4}} and {\ce{H2}}, FTIR transmittance spectra of each tube were recorded at three different temperatures: 300, 500, 700 K. Table~\ref{tab:exp_details_conditions} reports the experimental conditions in detail. 

\begin{table}[h]
\centering 
\caption{Experimental conditions and characteristics of the spectra.} \label{tab:exp_details_conditions}
\small
\begin{tabular}{ll} 
\toprule
 Parameter & Value \\
\hline
 Spectral coverage  	& 2800 -- 3200 cm$^{-1}$ \\ 
 Temperature range  	& 300 -- 700 K \\ 
 \ce{CH4} pressure 	& 0.8 -- 7.0 Torr \\ 
 \ce{H2} pressure      & 10.0 -- 933.3 Torr \\ 
\hline
 Cells path length  	        & 10.0$\pm$0.2 cm \\ 
 Number of averaged scans	& 100 -- 400\\ 
 Gas cell material/windows  	        & Quartz (\ce{SiO2})\\ 
 Gas cell transmission range   & 2750 -- 3250 cm$^{-1}$\\ 
 Light source  		& SiC Globar \\ 
 Beam splitter 		& KBr \\ 
 Detector 			& MCT LN2 \\ 
 Filter 			 & fused silica and Ge\\ 
 Spectral resolution (cm$^{-1}$) & 0.01 -- 0.005 \\ 
 Apodization function 	          & Box-car\\ 
\toprule
\toprule
\end{tabular}
\end{table} 

\begin{figure*} 
\centering
\includegraphics[scale=1.0]{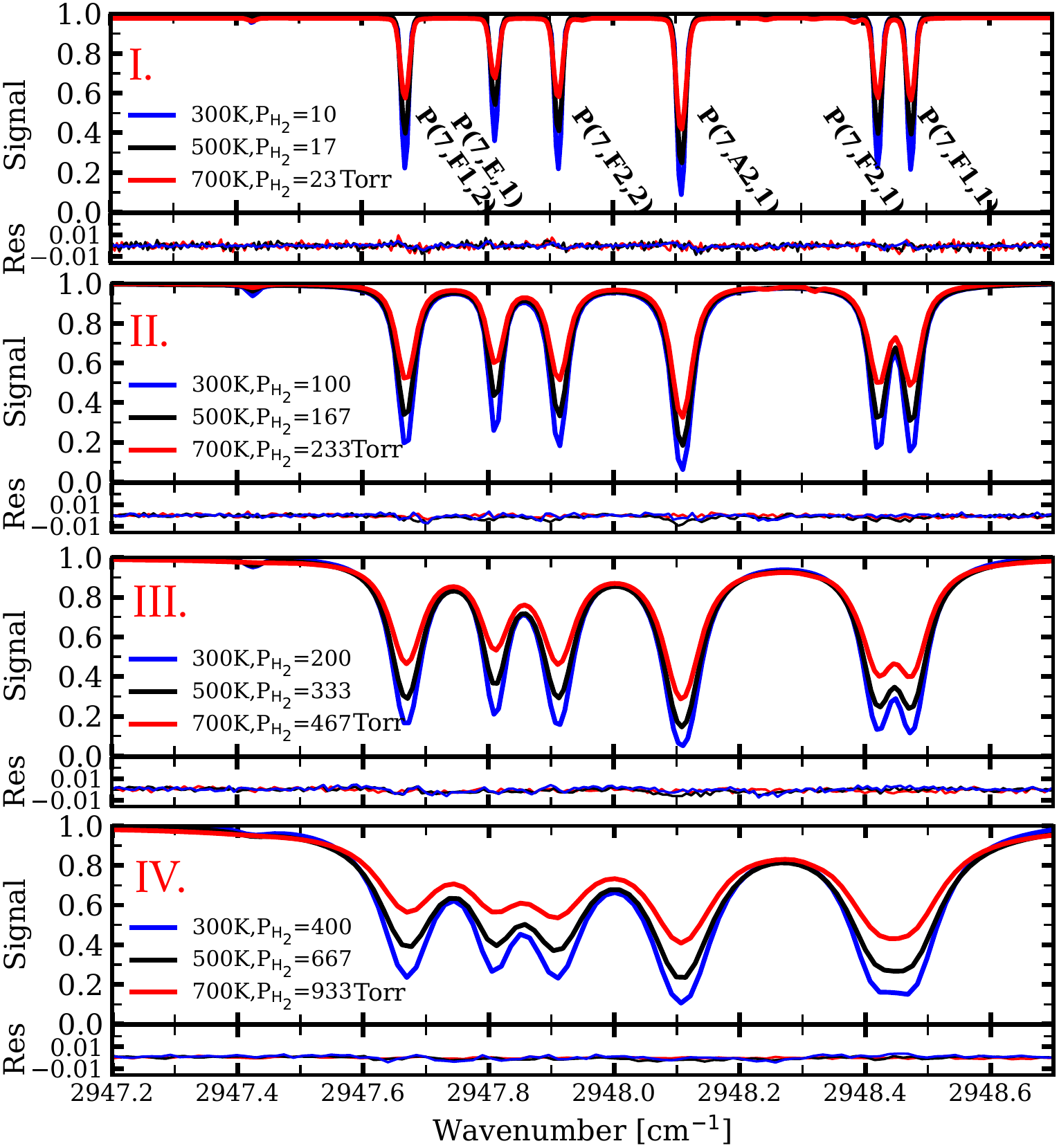} 
	\vspace{-0.2 cm}
        \caption{Examples of recorded spectra showing the $P(7)$ cluster of the $\nu_3$ band of \ce{CH4} in a \ce{H2} bath gas: In this study, four quartz cells (I--IV) with different amounts of P$_{\ce{CH4}}$ and P$_{\ce{H2}}$ (i.e., P$_{\ce{CH4}}$ (Torr) : P$_{\ce{H2}}$ (Torr) = 0.8:10, 1.1:100, 2.2:200, 3.0:400) were used to record the FTIR transmittance spectra of \ce{CH4} in \ce{H2} at 300 K (blue), 500 K (black), and 700 K (red). Transitions with $E$ symmetry are weaker than $A_1/A_2$ and $F_1/F_2$ symmetries for similar quantum number $J$. The measured resolution for these 12 spectra varies from 0.005 cm$^{-1}$ for the lowest pressure (panel I) to  0.02 cm$^{-1}$ for the highest pressure (panel IV, 700 K). A least-squares fitting procedure with Voigt line profiles was employed to model the spectra. The residual (Res) subpanels represent difference between modeled and recorded spectra. In addition, we also find lines from other bands such as $2 \nu_2$ and $\nu_2+\nu_4$ bands; however, their S/N is not strong enough for pressure-broadening analysis. }
        \label{fig:pb-signleJ}
\end{figure*}

Overall, 12 different spectra of \ce{CH4} were recorded at various resolutions. Figure~\ref{fig:Spectra-measured} represents an overview of spectrum $\#$4 (i.e., P$_{\ce{CH4}}$=1.1 Torr and P$_{\ce{H2}}$=100.0 Torr), which encompasses $P$, $Q$, and $R$ branches up to $J^{\prime\prime}$=17. In addition, each $J^{\prime\prime}$ consists of a cluster of transitions with various symmetry species and ${\it N}$ quantum index\footnote{ The \ce{CH4} energy levels are labelled by different quantum numbers such as $J$ and $C$ (tetrahedral symmetry), and ${\it N}$ (quantum index) defined in \citet{Brown1992}.}. Figure~\ref{fig:pb-signleJ} illustrates the modeled spectra for $P$(7) transitions.  The elevated temperature gas pressures, P$_{\ce{CH4}}$ and P$_{\ce{H2}}$, were then calculated using the ideal gas law. Table~\ref{tab: all spectra tubes} lists the resolution, number of scans, and the P$_{\ce{CH4}}$ and P$_{\ce{H2}}$ values for all measurements. Spectrum $\#$1 was used to measure the unbroadened Doppler-width and intensity of each line. 

\begin{table}[h]
\caption{Summary of the experimental conditions}\label{tab: all spectra tubes}
\vspace{0cm}
\centering 
\small
\begin{tabular}{p{5pt}p{13pt}p{15pt}p{25pt}p{20pt}p{30pt}p{30pt}} 
\toprule
$\#$ & Tube & $T$[K]& R$^{\color{red}\dagger}$[cm$^{-1}$] & Scan$^{\color{red}\ddagger}$ & $P_{\ce{CH4}}$[Torr]  & $P_{\ce{H2}}$[Torr]  \\
\hline
 1  & 1 & 300 & 0.005 & 400  & 0.8 & 10.0 \\ 
 2  & 1 & 500& 0.005 & 200  & 1.3 & 16.7  \\ 
 3  & 1 & 700 & 0.005 & 200  & 1.9 & 23.3 \\ 
\hline
 4 & 2 & 300 & 0.01 & 150 & 1.1 & 100.0  \\ 
 5 & 2 & 500 & 0.01 & 100 & 1.8 & 166.7  \\ 
 6 & 2 & 700 & 0.01 & 100 & 2.6 & 233.3  \\ 
\hline
 7  & 3 & 300 & 0.01 & 200 & 2.2 & 200.0  \\ 
 8  & 3 & 500 & 0.01 & 200 & 3.7 & 333.3  \\ 
 9  & 3 & 700 & 0.01 & 200 & 5.1 & 466.7  \\ 
\hline
 10  & 4 & 300 & 0.02 & 100 & 3.0 & 400.0  \\ 
 11  & 4 & 500 & 0.02 & 100 & 5.0 & 666.7  \\ 
 12  & 4 & 700 & 0.02 & 100 & 7.0 & 933.3  \\ 
\toprule
\toprule
\end{tabular}
\scriptsize
\hfill\parbox[t]{\linewidth}{Tube lengths are 10.3, 9.8, 10.2, and 9.9 cm, respectively. \\
  $\color{red}^{\dagger}$Spectral resolution.\\
  $\color{red}^{\ddagger}$Number of scans averaged.}

\end{table} 

The decomposition of \ce{CH4} is an important issue for high-temperature measurements \citep{Gueret1997}. To decrease the potential for loss of \ce{CH4}, we added 10 Torr of \ce{H2} into the first gas cell at room temperature. The main product of \ce{CH4 + CH4} bimolecular dissociation in the absence of \ce{H2} is \ce{CH3}, but in the presence of \ce{H2} gas as a third-body component, \ce{CH4} will reform. In other words, \ce{H2} gas will decrease the amount of decomposition by increasing the back reaction. Additionally, 10 Torr of \ce{H2} has a negligible pressure-broadening effect. { The volume mixing ratios of these gases can be calculated through minimizing Gibbs free energy which is dependent on the temperature, pressure, and gas concentrations. Therefore, we used the online thermodynamical simulator\footnote{http://navier.engr.colostate.edu/code/code-4/index.html} to calculate the fraction of decomposition of pure \ce{CH4} at different temperatures and pressures. Table~\ref{tab:ch4-mixratio} (case 1) represents the thermodynamic mixing ratios of 0.8 Torr of pure \ce{CH4}. Note that these calculations are done up to 900 K while the maximum laboratory temperature in this work is 700 K. Thermal decomposition of pure \ce{CH4} is predicted to occur for temperatures 700 K and above but is suppressed by the mixture of a small amount of \ce{H2}.} Ultimately, no significant decrease of the \ce{CH4} column density was noted even at 700 K. In this study, the line assignments and the line positions of \ce{CH4} were adopted from HITRAN2016 \cite{Gordon2017, Brown2013}.

\begin{table}[h]
\scriptsize 
\caption{Predicted thermodynamic \ce{CH4} volume mixing ratios{\color{red} $^\dagger$} }\label{tab:ch4-mixratio}
\vspace{-0.3cm}
\centering 
\begin{tabular}{cccccc} 
\toprule
  $T$(K)& \ce{CH4}($\%$) & \ce{H2}($\%$)  & \ce{C2H2}($\%$)   & \ce{C2H4}($\%$) \\
  \hline
  \multicolumn{6}{l}{Case 1: pure (100$\%$) 0.8 Torr of \ce{CH4} at 300 K{\color{red} $^\ddagger$}} \tabularnewline
  300 & 100    & 0.0    & 0.0              & 0.0        \\ 
  500 & 99.9   & 0.01   & 0.0              & 0.0        \\ 
  700 & 98.6   & 0.9    & 0.0              & 0.5        \\ 
  900 & 81.2   & 13.1   & 1.7              & 4.0       \\ 
\hline
  \multicolumn{6}{l}{Case 2: 0.8 (7.4$\%$) \ce{CH4} in 10.0 Torr (92.6$\%$) \ce{H2} at 300 K{\color{red} $^{\dagger\dagger}$} } \tabularnewline
  300 & 7.4   & 92.6    & 0.0        & 0.0    \\ 
  500 & 7.4   & 92.6    & 0.0        & 0.0    \\ 
  700 & 7.4   & 92.6    & 0.0        & 0.0    \\ 
  900 & 7.4   & 92.6    & 0.0        & 0.0    \\ 
\toprule
\toprule
 
\end{tabular}
\centering
\scriptsize
\hspace{2cm}\parbox[t]{\linewidth}{  $^\dagger$These mixing ratios are calculated by minimizing the Gibbs free energy of an ideal gas mixture.  \\
  $\color{red}^\ddagger$The reported pressure in each case is for 300 K. $P_{\ce{H2}}$ and $P_{\ce{CH4}}$ at high temperatures are calculated using ideal gas law.\\
 $\color{red}^{\dagger\dagger}$ In this study, 10 Torr of \ce{H2} gas added in order to suppress the \ce{CH4} decomposition.}
\end{table} 

The \ce{CH4} and \ce{H2} gases { were 99.99$\%$ } and obtained from Matheson. The gas pressure while filling the sample tubes was measured using two different MKS Baratron pressure gauges (maximum range 100 and 1000 Torr). For controlling the temperature, heat tapes from BriskHeat company (Type BW0) were used. A thermocouple connected to each gas cell was used in a feedback loop with the heat-tape controller to maintain a constant temperature. {  Omega company states typical uncertainties as 0.1$\%$ of the displayed reading for their digital readers.  The uncertainty for type K probes is estimated to be 0.75$\%$ (2.2 K at 300 K). Therefore, the overall uncertainty is due to the probe, not the reader, and  T is good to within $\pm$2 K at the location of the junction.} { There is a possibility of temperature nonuniformity in our gas cell. We expect this effect to be small given the high heating-element coverage of the cell excluding the transmitting windows but including its support structure, the small size of the cell, and its vacuum environment.} The uncertainty in the measurement of $P_{\ce{H2}}$ and { $P_{\ce{CH4}}$} is less than 0.5$\%$, and is also negligible.

\section{Data Analysis} \label{sec: analysis}
Our main goal is to extract pressure-induced broadening coefficients by modeling all lines with Voigt line profiles. Lorentzian and temperature-dependence coefficients for each rovibrational line are determined from linewidths extracted from spectra $\#$1--12 using a least-squares fitting method. The signal-to-noise ratio (S/N) is insufficiently high to justify modeling the spectra with non-Voigt profiles.

The negligibly pressure-broadened sample tube $\#$~1  was analyzed first to determine the correct \ce{CH4} line assignments, wavenumber calibration, and the presence of other \ce{CH4} bands and other contaminant species. Line strengths were determined separately at each measured temperature.  The highly-blended pressure-broadened spectra were analyzed with line strengths fixed to their unbroadened values and line widths and positions freely modified.

\subsection{Continuum / baseline fitting}
All \ce{CH4} spectra were converted from their interferograms with a  Boxcar apodization using the OPUS software\footnote{www.bruker.com}. The effect of instrumental broadening was modeled using a custom fitting code as a sinc function \cite{Bretzlaff1986}. The background continuum is also modeled using cubic splines optimised during the least-squares fitting procedure \cite{Burden2011}. Additionally, interference between the two cell windows that affects the recorded spectra by introducing sinusoidal behavior into the spectral continuum. We modeled this interference effect by employing two sine functions scaling the modeled spectrum.

\subsection{Line position corrections}
Line assignments are determined from  the recent version of HITRAN \cite{Brown2013, Gordon2017}. All corresponding line positions from HITRAN were input into the fitting code, and a global fit was made to calculate a single global shift induced by any slight miscalibration of the spectrometer. Afterward, the calculated shift was applied to our low pressure spectra (i.e., spectra $\#$1--3 in Table \ref{tab: all spectra tubes}). Later, the corrected/shifted line positions from the low pressure spectra were used to fit high pressure spectra (i.e., spectra $\#$4--12 in Table \ref{tab: all spectra tubes}), where pressure-induced lineshifts were also evident.

\subsection{Line profiles}	 
At very low pressure, the effect of collisions on molecular spectra is negligible. However, molecular velocities are distributed according to the Maxwell--Boltzmann statistics resulting in Doppler broadening (see chapter~1 at \cite{Buldyreva2011book}). The Doppler half-width at half-maximum (HWHM) linewidth (${\it \Gamma}_{\text{D}}$) were individually modeled using Gaussian line profile $f_{\text{G}}$ :

\begin{equation} \label{eq:doppler-profile}
     f_{\text{G}}(\nu - \nu_{ij}, {\it{\Gamma}}_{\text{D}}) = \sqrt{\frac{\ln(2)}{\pi{\it \Gamma}_{\text{D}}^2}}\exp\left(-\frac{\ln(2) (\nu - \nu_{ij})^2}{{\it \Gamma}_{\text{D}}^2}\right)
\end{equation}

\begin{equation} \label{eq:doppler-LW}
      {\it{\Gamma}}_{\text{D}}(T) = \frac{\nu_{ij}}{c}\sqrt{\frac{2 \ln(2) N_{\text{A}} k_{\text{B}} T}{M}}
 \end{equation}

 where $M$ is the molar mass of the absorber molecule in grams, $N_{\text{A}}$ is the Avogadro constant, $k_{\text{B}}$ is the Boltzmann constant, and $\nu_{ij}$ is the line position or the energy gap between quantum levels $i$ and $j$ in any arbitrary energy unit (e.g., cm$^{-1}$). ${\it \Gamma}_{\text{D}}$ values for our various measurements are in the range 0.004 -- 0.007 cm$^{-1}$ given the dependence of ${\it \Gamma}_{\text{D}}$ on the temperature and wavenumber. The natural radiative linewidth of the \ce{CH4} $\nu_3$ band is $\sim$ 10$^{-9}$ cm$^{-1}$ (i.e., in the range of 10--100 Hz) \citep{Jin1993}, which is fully negligible when fitting the spectra. 
 
Since the intensity of each line is distributed as a result of pressure-broadening, we increased the column density of \ce{CH4} when a high \ce{H2} pressure is present in order to obtain optimal S/N ratios without saturating any lines. As a result of this change, the modeled \ce{CH4} optical depth of high-pressure spectra (i.e., spectrum $\#$ 4--12) were scaled up uniformly.

The Lorentzian HWHM linewidth ${\it \Gamma}_{\text{L}}$ and lineshift ${\it \Gamma}_{\text{L}}$ were fitted individually for each line using the Lorentzian line profile $f_{\text{L}}$: 
	\begin{equation} \label{eq:loren-profile}
                f_{\text{L}}(\nu; \nu_{ij}, {\it\Gamma}_{\text{L}},{\it\Delta}_{\text{L}}) = \frac{1}{\pi}\frac{{\it\Gamma}_{\text{L}}(p,T)}{[{\it\Gamma}_{\text{L}}(p,T)]^2 + [\nu-(\nu_{ij} + {\it \Delta}_{\text{L}})]^2}
              \end{equation}

	\begin{equation} \label{eq:loren-HWHM}
		{\it \Gamma}_{\rm{L}}  =  \left( \frac{T}{T_0} \right) ^{-n_{\text{$\gamma$}}} \gamma_{\text{L}} P_{\ce{H2}}
	\end{equation}

	\begin{equation} \label{eq:loren-PB}
		{\it \Delta}_{\rm{L}}  =  \left( \frac{T}{T_0} \right) ^{-n_{\text{$\delta$}}} \delta_{\text{L}} P_{\ce{H2}}
	\end{equation}
 in which $\gamma_{\text{L}}$ (cm$^{-1}$/atm)  and n$_T$ are the Lorentzian linewidth coefficient and its temperature-dependence coefficient, respectively. $\delta_{\text{L}}$ (cm$^{-1}$/atm) and $n_{\text{$\delta$}}$ are the Lorentzian lineshift coefficient and its temperature-dependence coefficient, respectively. $T_{0}$ is a reference temperature, and it is set equal to 300 K. Note, all these coefficients are dependent on the total rotational quantum number of $J$, tetrahedral (T$_{\text{d}}$) symmetry species, and the broadeners. The code computes the Voigt profile as the Faddeeva function.\footnote{http://ab-initio.mit.edu/wiki/index.php/Faddeeva_Package}

 The Lorentzian coefficients ${\it \Gamma}_{\text{L}}$, extracted from the recorded spectra result from the effect of $P_{\rm{\ce{H2}}}$ collisional-induced broadening. The pressure-broadening from \ce{CH4} self broadening is negligible since $P_{\rm{\ce{CH4}}}$ $\leqslant $ 1.1$\% \times P_{\rm{\ce{H2}}}$. { Regarding Dicke narrowing, this effect becomes important at intermediate pressures or the Doppler--Lorentzian transition region because Doppler broadening at low pressures and Lorentzian broadening at high pressures mask the narrowing. For example, Pine \cite{Pine1992} found the largest discrepancy between Voigt and Rautian at 50 Torr \ce{H2}, and a corresponding 5$\%$ difference in the derived $\gamma_{\text{L}}$ for the two cases. This difference will be reduced by about half at 100 Torr (the lowest pressure we use). Then our Lorentzian linewidths fitted at 100 Torr may be underestimated by up to 3$\%$  (in comparison with random fitting uncertainties of at 4$\%$ or more).}

{  Other formulations for the temperature-dependence of Eq.~\ref{eq:loren-PB} have been adopted elsewhere \cite{Smith1992}. We use the most  conventional single-parameter temperature dependence formula above given the limited temperature sampling of our data. }

\section{Results and Discussion} \label{sec: results}

\subsection{Pressure broadening coefficients: ${\gamma}_{\rm L}$ and ${n}_{\gamma}$}
\begin{figure}[ht]
\centering
 \includegraphics[scale=0.8] {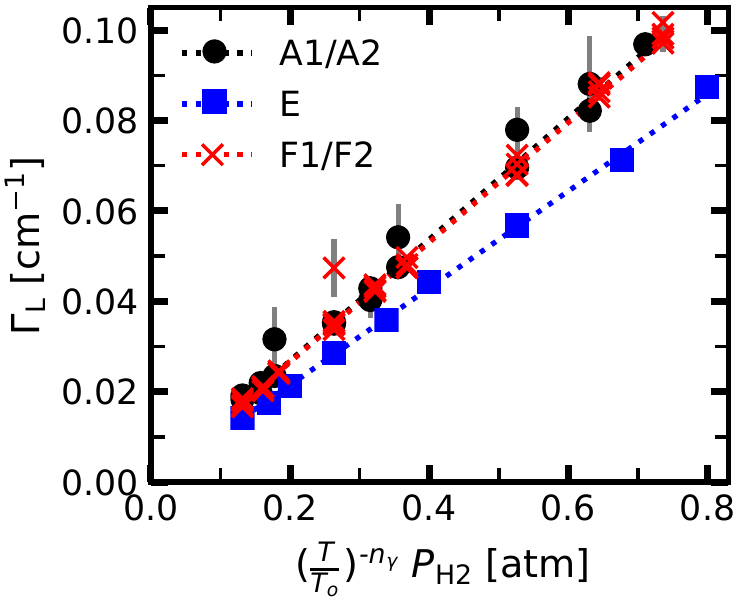} 
 \caption{Lorentzian linewidth ${\it \Gamma}_{\text{L}}$ versus (T/T$_0$)$^{-n_{\text{$\gamma$}}}$ P$_{\ce{H2}}$ for all transitions with $J^{\prime \prime}$=7: This plot shows that the Lorentzian coefficients are strongly dependent to the T$_{\text{d}}$ symmetry, and in most cases, $\gamma_{\text{L}}$ for $E$ symmetry is smaller than $A_1/A_2$ or $F_1/F_2$ symmetries.
   { Each point has its own error bar, which represents the uncertainly in fitting the ${\it \Gamma}_{\text{L}}$ linewidths. In addition, there are multiple lines with different $N$-index for the $A_1$/$A_2$ and $F_1/F_2$ symmetries.
Where error bars are not visible, the uncertainties are smaller than the symbol size.}}
	  \label{fig:PB-J7} 
\end{figure}

After fitting all 12 spectra from 300 to 700 K, the Lorentzian HWHM (i.e., ${\it \Gamma}_{\text{L}}$ in Eq.~\ref{eq:loren-HWHM}) is extracted for each tetrahedral rovibrational transition\footnote{{ Each tetrahedral transition is labelled by total rotational quantum number $J$, symmetry species $C$, and quantum index $N$ \cite{Brown1992}.}}.
Then, the $\gamma_{\text{L}}$ and $n_{\text{$\gamma$}}$ coefficients are computed in three different ways: 1) for all lines individually including its own $J$, symmetry and $N$ numbers, 2) averaged over lines with the same $J^{\prime\prime}$ but different symmetry and $N$ index (i.e., the multiplicity index), and 3) all lines with the same $J^{\prime\prime}$ and symmetry but different $N$ index were fitted. As a sample fitting, Fig.~\ref{fig:PB-J7} illustrates ${\it \Gamma}_{\text{L}}$ versus (T/300 K)$^{-n_{\text{$\gamma$}}}$$P_{\ce{H2}}$ for $J^{\prime\prime}$=7 and different symmetry species ($n_{\text{$\gamma$}}$ is computed below). Figure~\ref{fig:PB-J7} shows the fitted slope (i.e., $\gamma_{\text{L}}$) of transitions with $A_1$/$A_2$ and $F_1$/$F_2$ is higher than for the $E$ symmetry lines.

\begin{figure*}[h!] 
\centering
 \includegraphics[scale=0.9] {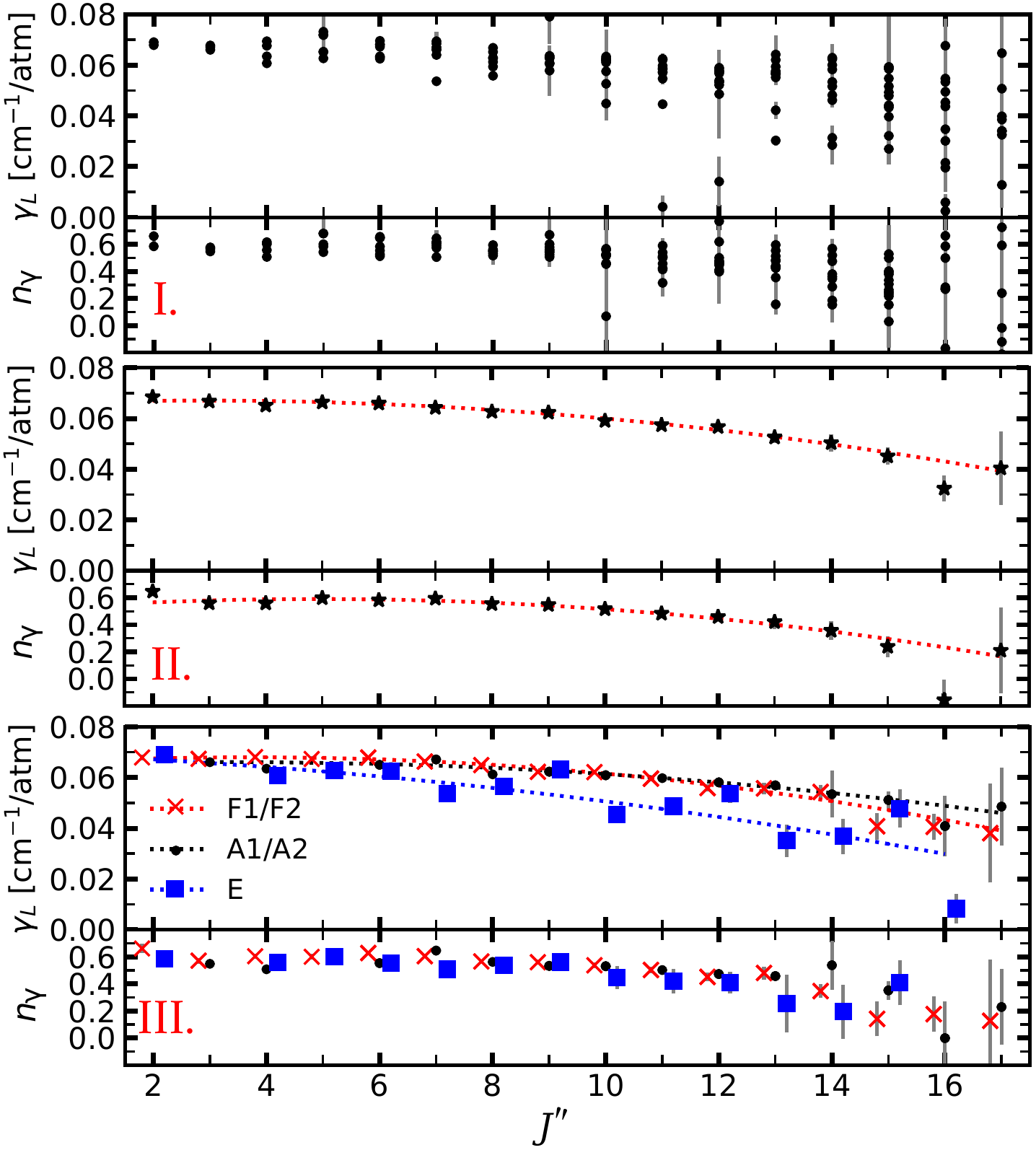}
 \caption{Dependence of the Lorentzian coefficients $\gamma_{\text{L}}$ and $n_{\text{$\gamma$}}$ on quantum number $J^{\prime\prime}$.  Panel I: All lines were fitted individually. The uncertainties are 1-$\sigma$ error in each individual line but the scatter arises primarily from the symmetry and ${\it N}$-index. Panel II: Average of all lines over symmetry and $N$ index. However this approach ignores the dependence of symmetry and ${\it N}$ dependency, and therefore, it disregards these important physical effects. Panel III: Average of all lines within a symmetry species. $\gamma_{\text{L}}$ and $n_{\text{$\gamma$}}$ are reported in Tables \ref{tab:PB_ALL_J_Dependence}, \ref{tab:PB_ALL_J_SYM_Dependence}, and in the Table S1 (supplementary file). Note that a few points are out of the fitted trend (dotted line), and therefore, the fitted coefficients are reported in these tables. In addition, weak lines with low S/N ratio and high-uncertainty are removed from the list in Table S1 (supplementary file). { Where error bars are not visible, the uncertainty from fittings for the data is smaller than the symbol size itself.}}
 \label{fig: pb-all plots}
\end{figure*}

Figure~\ref{fig: pb-all plots}(I--III) illustrates the trend of $\gamma_{\text{L}}$ and $n_{\text{$\gamma$}}$ with $J^{\prime\prime}$. Figure \ref{fig: pb-all plots}(I) represents $\gamma_{\text{L}}$ and $n_{\text{$\gamma$}}$ fitted to all lines individually. At each $J^{\prime\prime}$ value, there is the scatter of both $\gamma_{\text{L}}$ and $n_{\text{$\gamma$}}$ coefficients which arise from the difference between T$_{\text{d}}$ symmetries, $N$ indexes, and random fitting errors. { In the first analysis step, individual lines with the same $J, symmetry, N$ from all spectra were fitted to extract the $\gamma_{\text{L}}$ and $n_{\text{$\gamma$}}$ coefficients data. From this we determine the Lorentzian linewidth of each individual line as a result of \ce{H2} collisional impact. Figures \ref{fig:PB-J7} and \ref{fig: pb-all plots}(I) as well as the supplementary Table (S1) represent these results. The error bars shown in these figures and the table uncertainties are due to the fitting uncertainties, noise, and the low signal-to-noise of some lines. These line-by-line coefficients are the main outcome of this study and they can be utilized in generating absorption cross-section (or opacity) data the standard HITRAN code\footnote{i.e. HITRAN Application Programming Interface (HAPI) \citep{Kochanov2016}, https://github.com/hitranonline/hapi} or the NASA Ames Freedman's code \citep{Freedman2014,Freedman2008}.}

In contrast, if we average the coefficients for all lines with the same $J^{\prime\prime}$ value, then $\gamma_{\text{L}}$ and $n_{\text{$\gamma$}}$ coefficients fall in the range of 0.07--0.03  and 0.65--0.25, respectively (see Fig.~\ref{fig: pb-all plots}(II) and Table \ref{tab:PB_ALL_J_Dependence}). In Table~\ref{tab:PB_ALL_J_Dependence}, the scatter of these coefficients are mostly due to the scattering of lines with the same $J^{\prime\prime}$ but different symmetries and ${\it N}$ dependencies, as well as, the uncertainty in fitting the Lorentzian linewidths from the recorded spectra.
{ Another motivation for this step is to provide data for opacity codes which input only $J$-dependent pressure-broadening values such as the current version of EXOCROSS code\footnote{https://github.com/Trovemaster/exocross}\citep{Yurchenko2018}. Figure \ref{fig: pb-all plots}(II) shows that there is a clear dependency of the Lorentzian coefficient and its temperature-dependence with ($\gamma_{\text{L}}$ and $n_{\text{$\gamma$}}$) on $J$. This data are also presented in Table~\ref{tab:PB_ALL_J_Dependence}, and the range of scatter for each one is shown as a range of $\gamma_L^{min}$--$\gamma_L^{max}$ and $n_{\text{$\gamma$}}^{min}$--$n_{\text{$\gamma$}}^{max}$.} According to the Anderson collisional theory \citep{Anderson1949}, the $n_{\text{$\gamma$}}$ coefficient is expected to be 0.5; however, our analysis shows that $n_{\text{$\gamma$}}$ coefficients deviate from this value by up to 30$\%$. We also find that $\gamma_{\text{L}}$ and $n_{\text{$\gamma$}}$ decrease by 25$\%$ and 80$\%$ , respectively, between $J ^{\prime \prime }$=2 and 17 in agreement with the trend calculated by \citet{Neshyba1994} and \citet{Gabard2013}.

{ Next, we grouped the lines with similar symmetries, and extracted the Lorentzian coefficients from each group. Figure \ref{fig: pb-all plots}(III) shows the symmetry-dependence of $\gamma_{\text{L}}$ and $n_{\text{$\gamma$}}$. The bars shown in this figure are due to the uncertainty in fitting this data (similar to Fig.~\ref{fig: pb-all plots}(III)) and also the scatter imposed by different values of the $N$ quantum index. It should be noted that only some symmetry-$J$ combinations have  multiple $N$ values. Therefore, two kinds of uncertainties are shown in Table~\ref{tab:PB_ALL_J_SYM_Dependence}: statistical fitting uncertainties for singular-$N$ values, and the range of scatter for values averaged over mutiple $N$ transitions.} { In general, within each $J$ manifold, $E$-lines are the weakest and also have the narrowest Lorentzian linewidth ${\it \Gamma}_{\text{L}}$. In contrast, lines with ${A_1/A_2}$ and ${F_1/F_2}$ symmetries are generally the strongest, and have the broadest linewidth. }

{ Following the complex Robert-Bonamy theory\cite{RobertBonamy1979}, \citet{Neshyba1994} calculated the impact of electrostatic and atom-atom intermolecular potential on the line broadening and line shift of the \ce{CH4}@[\ce{N2}] system. They found that the atom--atom potential component is the main reason for the line broadening with a corresponding decrease with increasing total angular momentum, $J$. In addition, the  broadening effect is symmetry dependent and it  was shown \citep{Smith1981, Smith1992, Gabard2013} the total collisional cross-section for $E$ symmetry is lower than for  $A_1/A_2$ and $F_1/F_2$ at low $J$, which results in smaller perturbation and collisional-broadening for the $E$-symmetry species, as we observed.}). 

\begin{figure}[h] 
\centering
 \includegraphics[scale=0.9] {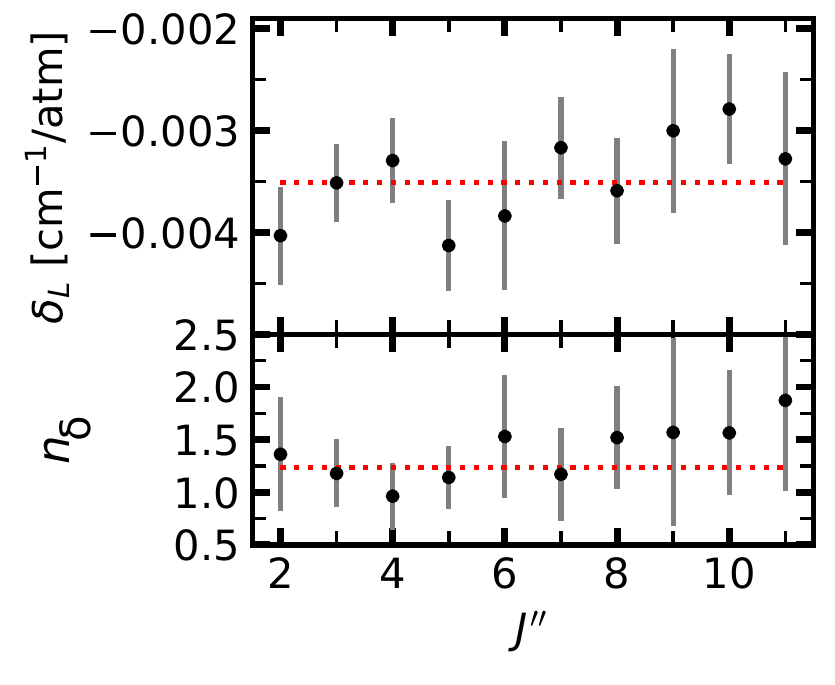} 
	  \caption{ The dependency of pressure shift on quantum number $J^{\prime\prime}$. The plotted error-bars are an indistinguishable combination of fitting uncertainties and differences between symmetry and $N$ quantum index.  } \label{fig:ps-data}
\end{figure}

\begin{table}[h]
  \caption{$J$-dependent Lorentzian coefficients averaged over $\nu_3$ $P$ branch line cluster of \ce{CH4}.}
  \vspace{-0.2cm}
  \label {tab:PB_ALL_J_Dependence}
  \centering
  \footnotesize
  \begin{tabular}{p{1pt}p{1pt}p{11pt}p{40pt}p{6pt}p{39pt}p{26pt}p{10pt}}
\toprule
$J^{\prime}$ &  $J^{\prime \prime}$  & $\gamma_{\text{L}}$ & $\gamma_L^{min}$--$\gamma_L^{max}{\color{red}^\star}$ & $n_{\text{$\gamma$}}$ & $n_{\text{$\gamma$}}^{min}$--$n_{\text{$\gamma$}}^{max}{\color{red}^\star}$  & $\delta_{\text{L}}$${\color{red}^\dagger}$ & $n_{\text{$\delta$}}$ \\
\hline
1&2&   0.069&  0.068--0.069  & 0.65&  0.59--0.66  &\\
2&3&   0.067&  0.066--0.068  & 0.56&  0.55--0.57  &$-$0.0040(5) &  1.4(5) \\
3&4&   0.065&  0.061--0.068  & 0.56&  0.51--0.60  &$-$0.0035(4) &  1.2(3) \\
4&5&   0.066&  0.063--0.067  & 0.60&  0.54--0.68  &$-$0.0032(4) &  1.0(3) \\
5&6&   0.066&  0.063--0.068  & 0.58&  0.55--0.63  &$-$0.0041(4) &  1.1(3) \\
6&7&   0.064&  0.054--0.067  & 0.59&  0.51--0.65  &$-$0.0038(7) &  1.5(6) \\
7&8&   0.063&  0.057--0.061  & 0.55&  0.54--0.57  &$-$0.0031(5) &  1.2(4) \\
8&9&   0.063&  0.062--0.063  & 0.55&  0.53--0.56  &$-$0.0035(5) &  1.5(5) \\
9&10&  0.059&  0.045--0.062  & 0.52&  0.44--0.54  &$-$0.0030(8) &  1.6(9) \\
10&11& 0.058&  0.049--0.060  & 0.48&  0.42--0.52  &$-$0.0027(5) &  1.6(6) \\
11&12& 0.057&  0.054--0.058  & 0.46&  0.41--0.47  &$-$0.0032(8) &  1.9(9) \\
12&13& 0.053&  0.035--0.057  & 0.42&  0.25--0.48  &&\\
13&14& 0.051&  0.037--0.054  & 0.36&  0.20--0.54  &&\\
14&15& 0.046&  0.041--0.051  & 0.24&  0.14--0.41  &&\\
15&16& 0.043${\color{red}^\ddagger}$& &0.24${\color{red}^\ddagger}$ &  & & \\
16&17& 0.041&  0.038--0.049  & 0.20& 0.14--0.65& & \\
\toprule
\toprule
\end{tabular}
 \scriptsize
 \parbox[t]{\linewidth}{Note:\\
   ${\color{red}^\star}$ $\gamma_L^{min}$--$\gamma_L^{max}$ and $n_{\text{$\gamma$}}^{min}$--$n_{\text{$\gamma$}}^{max}$ represent the range of coefficients before averaging over symmetry and ${\it N}$. Only the lines with high S/N ratios are considered for extracting the Lorentzian pressure-shift coefficients ($\delta_{\text{L}}$ and $n_{\text{$\delta$}}$).\\
   ${\color{red}^\dagger}$The scattering in the pressure-shift coefficients arises from their dependencies into the symmetry and ${\it N}$. \\
 ${\color{red}^\ddagger}$The extracted values of  $\gamma_{\text{L}}$ and $n_{\text{$\gamma$}}$ for $J^{\prime \prime}$=16 are 0.0324(51) and -0.2(1), which are out of the trend. Therefore, these values are replaced with the expected values from the polynomial equation~\ref{eq:fit-poly-J} due to the weakness of the lines.   }
\end{table}

\subsection{ Lorentzian temperature-dependence coefficient: $n_{\text{$\gamma$}}$} \label{sec:n-gamma-4-2}
According to early Anderson collisional theory \cite{Anderson1949, Baranger1958} a broadened line has a Lorentzian profile (Eq.~\ref{eq:loren-profile}), and the broadening linewidth is proportional to T$^{-0.5}$ following Eqs.~\ref{eq:anderson_gamma} and \ref{eq:anderson-gamma-2}:
\begin{equation}
  \label{eq:anderson_gamma}
  {\it \Gamma}_{\rm L} = \frac{n \bar{\nu}_{\rm th} \sigma_r}{2\pi}
\end{equation}
where $n$ is the broadener column density (i.e., $n$=$n_{\ce{H2}}$= $P_{\ce{H2}}/k_bT$), $\bar{\nu}_{\rm th}$ is the mean thermal velocity from Maxwell-Boltzmann distribution (i.e., $\bar{\nu}_{\rm th}=\sqrt{8k_bT/\pi m}$ where $m$ is the \ce{H2} mass), and $\sigma_r$ is the real component of the collisional cross-section (see discussion  in \cite{Lyons2018}). 
\begin{equation}
  \label{eq:anderson-gamma-2}
  {\it \Gamma}_{\rm L} = \sqrt{\frac{2}{k_{\rm B} m \pi^3}} { P_{\ce{H2}}}T^{-0.5} \sigma_r 
\end{equation}

Following Eqs. \ref{eq:anderson_gamma} $\&$ \ref{eq:anderson-gamma-2}, the temperature-dependence coefficient, $n_{\text{$\gamma$}}$ is 0.5. Note, there are different assumptions at play in Eq.~\ref{eq:anderson-gamma-2} including the hard-sphere approximation, ideal gas law, and also a single thermal velocity $\bar{\nu}_{th}$ for all broadeners. { Therefore, this $n_{\text{$\gamma$}}$=0.5 value should be considered as a gas kinetic value, and a more sophisticated picture is reviewed by \citet{Gamache2018}}. Our results show that $n_{\text{$\gamma$}}$ strongly depends on $J$, and it is in the range of $\sim$0.65--0.2 (see Fig.~\ref{fig: pb-all plots}(II)). No significant dependence of $n_{\text{$\gamma$}}$ on the tetrahedral symmetry species is found.

Table \ref{tab:Literature-nT} lists most previous temperature-dependence measurements of \ce{CH4} in different broadeners. In addition, the measurements are for different fundamental and combination vibrational modes providing insight into the vibrational dependency of $n_{\text{$\gamma$}}$. The $n_{\text{$\gamma$}}$ of \ce{CH4}@[\ce{N2}] and \ce{CH4}@[Air] falls in the range of $\sim$0.55--1.0 and $\sim$0.4--0.9, respectively, which are roughly 30$\%$ larger than our results for \ce{CH4}@[\ce{H2}]. In comparison, \ce{CH4}@[\ce{He}] is about half that of \ce{CH4}@[\ce{H2}]. Table~\ref{tab:Literature-nT} also illustrates the slight vibrational-dependency of $n_{\text{$\gamma$}}$ for various broadeners.

\begin{table}[h] 
\footnotesize 
  \caption{Overview of previous measured temperature-dependence coefficients, $n_{\text{$\gamma$}}$, of \ce{CH4} in various broadeners}\label{tab:Literature-nT}
\centering 
\begin{tabular}{p{24pt}p{35pt}p{24pt}p{14pt}p{35pt}p{24pt}}
\toprule
Broadener& T[K]& Band   & Lines & $n_{\text{$\gamma$}}$$^{\color{red}\dagger}$ & Ref.\\
\hline
 {\bf \ce{H2}}  & {\bf 300$\ $ -- 700} & {\bf $\nu_3$}   & {\bf 116} & {\bf 0.2 -- 0.65 }& PS$^{\color{red}\ddagger}$\\
& 77$\ $ -- 295 & 6$\nu_1$,5$\nu_3$& 2  & 0.45, 0.53   &\cite{Keffer1986} \\ 
& 130 -- 295& $\nu_4$   & 6  & 0.46 -- 0.51 &\cite{Varanasi1989CH4nT,Varanasi1989ch4_v4} \\ 
& 161 -- 295& $\nu_4$   & 6  & 0.35 -- 0.52&\cite{Varanasi1990} \\ 
\hline
 Air  & 200 -- 300 & $\nu_3$&  3 & 0.62 -- 1.0$\ $  & \cite{Varanasi1975} \\ 
& 211 -- 314& $\nu_4$   & 148& 0.50 -- 0.80 & \cite{Smith1992}   \\ 
& 212 -- 297& $\nu_1+\nu_4$ & 130& 0.50 -- 0.85 & \cite{Devi1994}\\ 
& 212 -- 297& $\nu_3+\nu_4$ & 406& 0.50 -- 0.90 & \cite{Devi1994}\\ 
& 212 -- 297& $\nu_2+\nu_3$ & 71 & 0.40 -- 0.85 & \cite{Devi1994}\\ 
\hline
 \ce{N2}& 215 -- 297& $\nu_3$   &  3 & 0.94 -- 0.97 &\cite{Devi1983} \\ 
& 215 -- 297& $\nu_2+\nu_4$ &  2 & 0.86,$\ $ 0.92   &\cite{Devi1983} \\ 
&  \hspace{1pt} 77 $\ $-- 295  & 6$\nu_1$,5$\nu_3$  &  2 & 0.77,$\ $ 0.97   &\cite{Keffer1986} \\ 
& 130 -- 295& $\nu_4$   &  6 & 0.75 -- 0.83 &\cite{Varanasi1989CH4nT,Varanasi1989ch4_v4}\\ 
& 161 -- 295& $\nu_4$   &  6 & 0.71 -- 0.82 &\cite{Varanasi1990} \\ 
& 211 -- 314& $\nu_4$   & 148& 0.55 -- 0.85 &\cite{Smith1992}\\ 
&  \hspace{1pt} 90 $\ $-- 296& $\nu_3$  & 4  & 0.84 -- 0.86 &\cite{Mondelain2007} \\ 
\hline
  Self   &77$\ $ -- 295 & 6$\nu_1$,5$\nu_3$  & 2  & 0.84, 0.93  &\cite{Keffer1986} \\
\hline
 \ce{He}&  \hspace{1pt} 77 -- 295$\ $ & 6$\nu_1$,5$\nu_3$& 2 & 0.37, 0.67   &\cite{Keffer1986} \\ 
& 130 -- 295& $\nu_4$   & 6  & 0.28 -- 0.38 &\cite{Varanasi1989CH4nT,Varanasi1989ch4_v4} \\ 
& 161 -- 295& $\nu_4$   & 6  & 0.26 -- 0.38  &\cite{Varanasi1990} \\ 
\hline
 \ce{Ar}& 130 -- 295& $\nu_4$   & 2  & 0.80 -- 0.83 &\cite{Varanasi1989CH4nT,Varanasi1989ch4_v4} \\ 
& 161 -- 295& $\nu_4$   & 6  & 0.72 -- 0.82  &\cite{Varanasi1990} \\ 
\toprule
\toprule

\end{tabular}
 \scriptsize
 \hfill\parbox[t]{\linewidth}{
   $\color{red}^{\dagger}$ $n_{\text{$\gamma$}}$ coefficients are reported within a range due to their dependency on $J$ and vibrational quantum numbers. In this table, the reported $n_{\text{$\gamma$}}$ in some cases are extracted from a few transitions. \\
   $\color{red}^{\ddagger}$ Present Study\\ }

\end{table} 

\subsection{ Lorentzian line-shift coefficients: $\delta_{\text{L}}$ and $n_{\text{$\delta$}}$}

The S/N ratio in the current study is  insufficiently high to extract pressure shifts for all lines. Hence, a pressure-shift coefficient is calculated only for lines with $J^{\prime\prime}$=2--11, and falls in the range of $-0.004$ to $-0.002$ cm$^{-1}$/atm. We discern no significant dependence of the Lorentzian pressure-shift coefficients on the $J$ values (Fig.~\ref{fig:ps-data}), and the mean value of $\delta_{\text{L}}$ and $n_{\text{$\delta$}}$ are $-$0.0035 cm$^{-1}$/atm and  1.24, respectively. { The reported uncertainties for Lorentzian pressure-shift coefficients ($\delta_{\text{L}}$ and $n_{\text{$\delta$}}$) are due to the scatter  of the symmetry- and ${\it N}$-dependency.} Note that our $n_{\text{$\delta$}}$ is larger than $n_{\text{$\gamma$}}$, and this difference has been reported for water self-broadening as well \citep{Markov1994}.

{ The form of Eq.~\ref{eq:loren-PB} is based on Eq.~\ref{eq:loren-HWHM}, which is derived from the ideal gas law and hard-sphere approximation. Some studies of other systems such as \citet{Frost1976} and \citet{Baldacchini1996} have shown temperature-dependence has more complex form than our selected formula in Eq.~\ref{eq:loren-PB}. Additionally, \citet{Smith1992} found both positive and negative $\delta_{\text{L}}$ and $n_{\text{$\delta$}}$ values for the \ce{CH4} $\nu4$ band. However, we exclusively observed negative $\delta_{\text{L}}$ and positive $n_{\text{$\delta$}}$ values.}

\subsection{ Global equations for Lorentzian coefficients}
In order to provide Lorentzian broadening coefficients ($\gamma_{\text{L}}$ and $n_{\text{$\gamma$}}$) appropriate  for high-temperature \ce{H2}-dominanted exoatmospheres (i.e., super-Earth or warm-Neptunes with 400--900 K temperature), the dependency of these coefficients with $J^{\prime\prime}$ is presented by fitting the experimental results to a second-order polynomial $J^{\prime\prime}$-dependence (e.g., Eq.~\ref{eq:fit-poly-J}  see the red-dashed line in Fig.~\ref{fig: pb-all plots}(II)). Additionally, due to the significant dependence of $\gamma_{\text{L}}$ on the symmetry species, the fitting coefficients are extracted from them separately (i.e., Eq.~\ref{eq:fit-poly-J-sym}  dashed lines in Fig.~\ref{fig: pb-all plots}(III)) conforming to:
\begin{equation}
  \label{eq:fit-poly-J}
X(J^{\prime\prime})=m_1 + m_2 J^{\prime\prime} + m_3 {J^{\prime\prime}}^2
\end{equation}
\vspace{-0.5cm}
\begin{equation}
  \label{eq:fit-poly-J-sym}
X(J^{\prime\prime}, {\rm sym})=m_1 + m_2 J^{\prime\prime} + m_3 {J^{\prime\prime}}^2
\end{equation}

where $m_1$, $m_2$, and $m_3$ are the fitted constants, $X$ is the Lorentzian coefficient i.e., $\gamma_{\text{L}}$, $n_{\text{$\gamma$}}$,  and ``sym'' can be $A_1/A_2$, $F_1/F_2$, or $E$ symmetry species.
All the polynomial fitted constants are presented in Table~\ref{tab:polynominal-m}.

\begin{table}[H]
  \caption{Fitted constants for global equations${\color{red}^{\dagger}}$}
  \vspace{-0.2cm}
  \label {tab:polynominal-m}
\centering
\small
\begin{tabular}{ l c c c }

\toprule
Case &  $m_1$ &  $m_2$  & $m_3$ \\
\hline
$\gamma_{\rm {L}}$ ($J^{\prime\prime}$)      &  0.066 & $\ $0.0008 & $-$0.00014  \\
  $n_{\text{$\gamma$}}$ ($J^{\prime\prime}$)         &  0.520 & $\ $0.0290 & $-$0.00290 \\
  \hline
$\gamma_L$ ($J^{\prime\prime}$, $F_1/F_2$) & 0.0657 &      $\ $0.0012 &  $-$0.00017\\
$\gamma_L$ ($J^{\prime\prime}$, $A_1/A_2$) & 0.0650  &     $\ $0.0007 &  $-$0.00011\\
$\gamma_L$ ($J^{\prime\prime}$, $E$)       & 0.0690  &  $-$0.0010 &  $-$0.00010\\
\toprule
\toprule
\end{tabular}
 \footnotesize
 \hfill\parbox[t]{\linewidth}{
   ${\color{red}^{\dagger}}$Eqs.~\ref{eq:fit-poly-J} and \ref{eq:fit-poly-J-sym} are used to fit the $\gamma_{\text{L}}$ and $n_{\text{$\gamma$}}$ results.\\
     Note: $\gamma_{\text{L}}$ and $n_{\text{$\gamma$}}$ coefficients are reported in Table S1 for all $\nu_3 \ P$-branch lines. The polynomial equations Eqs.~\ref{eq:fit-poly-J} and \ref{eq:fit-poly-J-sym} should be used to determine these coefficients for transitions in the range of $J^{\prime\prime}$=0--18.   }
\end{table}

\subsection{Comparison with existing data}
Since \ce{CH4} is an important molecule in the atmosphere of the Earth, other planets, and brown dwarfs, many experiments have been carried out  for broadeners in the atmosphere of Earth (i.e., \ce{N2} and \ce{O2}), Jupiter (i.e., \ce{H2} and \ce{He}), and other broadeners such as Ar- and self-broadening. In the following, we will discuss the comparison of our results with the most relevant literature data.

\begin{figure*}[h] 
\centering
 \includegraphics[scale=1.05] {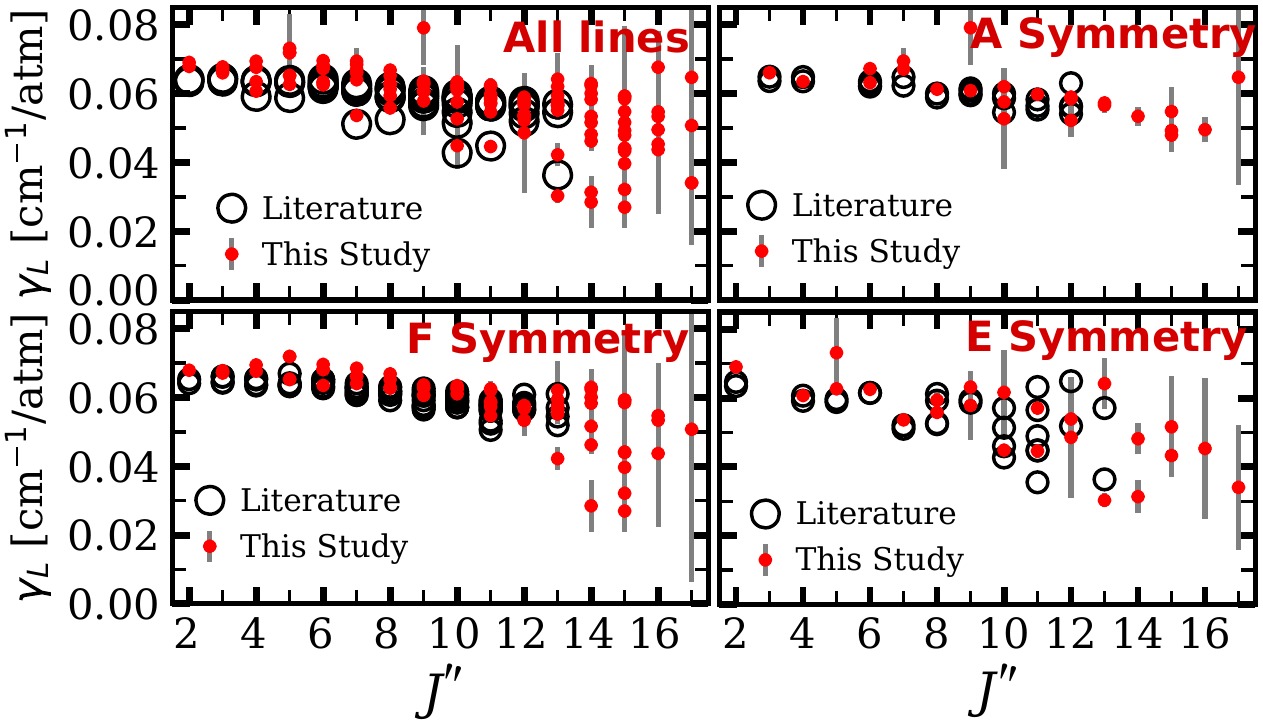} 
 \caption{Comparison of our $\gamma_L$ results with literature pressure-broadening coefficients of
   $\nu_3 \ Q$-branch \ce{CH4}@[\ce{H2}] \cite{Pine2019_CH4_v3_theory_nonVoigt, Pine1992}, and 
   $\nu_3 \ P$-branch \ce{CH4}@[\ce{H2}] \cite{Es-sebbar2014_CH4_P_v3}. { Where error bars are not visible, the uncertainties for our $\gamma_{\text{L}}$ are smaller than the symbol size itself.}}
 \label{fig:compare-pb-symmetry}
\end{figure*}

Figure \ref{fig:compare-pb-symmetry} represents the comparison of our results with the literature data \cite{Pine2019_CH4_v3_theory_nonVoigt, Pine1992, Varanasi1989CH4nT,Varanasi1989ch4_v4, Es-sebbar2014_CH4_P_v3} for the \ce{CH4}@[\ce{H2}] $\nu_3$ band. Note that most of the previous studies have been for the ${\it Q}$ branch \cite{Pine2019_CH4_v3_theory_nonVoigt, Pine1992} and employed  Rautian line profiles; while there are a few measurements on the ${\it P}$ branch \cite{Varanasi1989CH4nT, Es-sebbar2014_CH4_P_v3}, none employ \ce{H2} as a broadener. Figure \ref{fig:compare-pb-symmetry} (I,II) shows the comparison of all lines with their $J$-, symmetry- and ${\it N}$-dependencies.
In Fig.~\ref{fig:compare-pb-symmetry} (II), the \citet{Pine2019_CH4_v3_theory_nonVoigt} results are slightly lower than ours which might be due to the selection of different line profiles and branches.
In Fig. \ref{fig:compare-pb-symmetry}, a comparison of lines within different symmetry classes is shown.        

\begin{figure*}[h] 
\centering
 \includegraphics[scale=1.0] {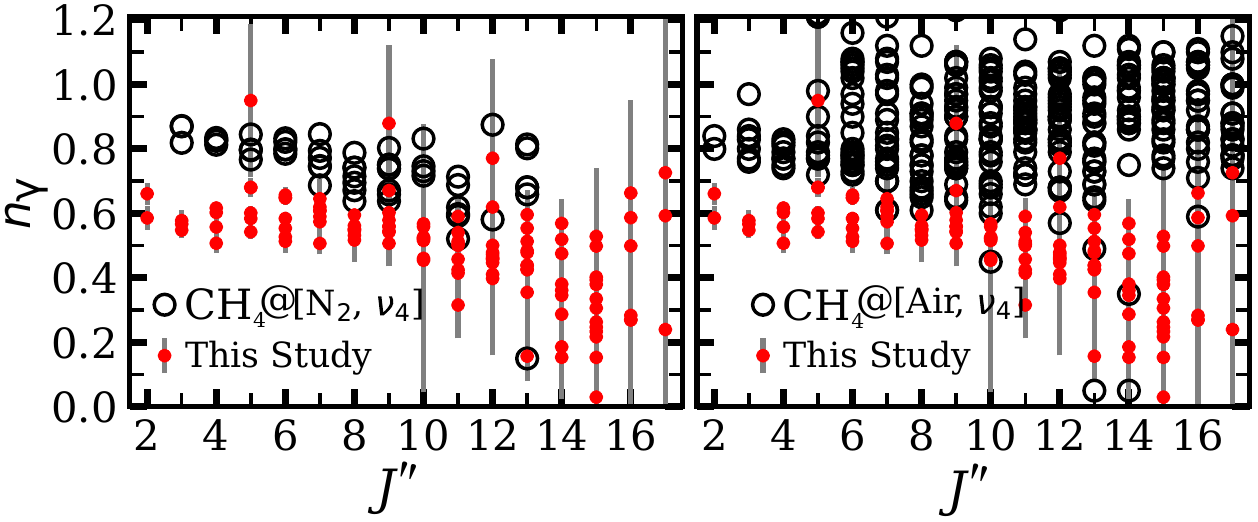} 
 \caption{Comparison of our temperature-dependence coefficients (i.e., $\nu_3 \ P$-branch \ce{CH4}@[\ce{H2}]) at temperature range 300--700 K with
   $\nu_4 \ Q$-branch \ce{CH4}@[\ce{N2}] \citet{Smith1992}  and
   $\nu_4 \ P$- and $R$-branch \ce{CH4}@[Air] \citet{Smith2009} at temperature range of 210--314 K.} \label{fig:compare-nT}
\end{figure*}

Figure~\ref{fig:compare-nT} illustrates the comparison between our temperature-dependence coefficients $n_{\text{$\gamma$}}$ (i.e., \ce{CH4}@[\ce{H2}] for $\nu_3$ $P$ branch) with both \ce{CH4}@[\ce{N2}] and \ce{CH4}@[Air] for $\nu_4$ band. Note, there are a two differences between these measurements: 1) our broadener \ce{H2} is different from the previous works, 2) there might be some vibrational-dependency of $n_{\text{$\gamma$}}$. In general our $n_{\text{$\gamma$}}$ coefficients (@[\ce{H2}]) is smaller than both @[Air]-broadening and @[\ce{N2}]-broadening \ref{tab:Literature-nT}.

\begin{figure*}[h] 
\centering
\includegraphics[scale=0.8] {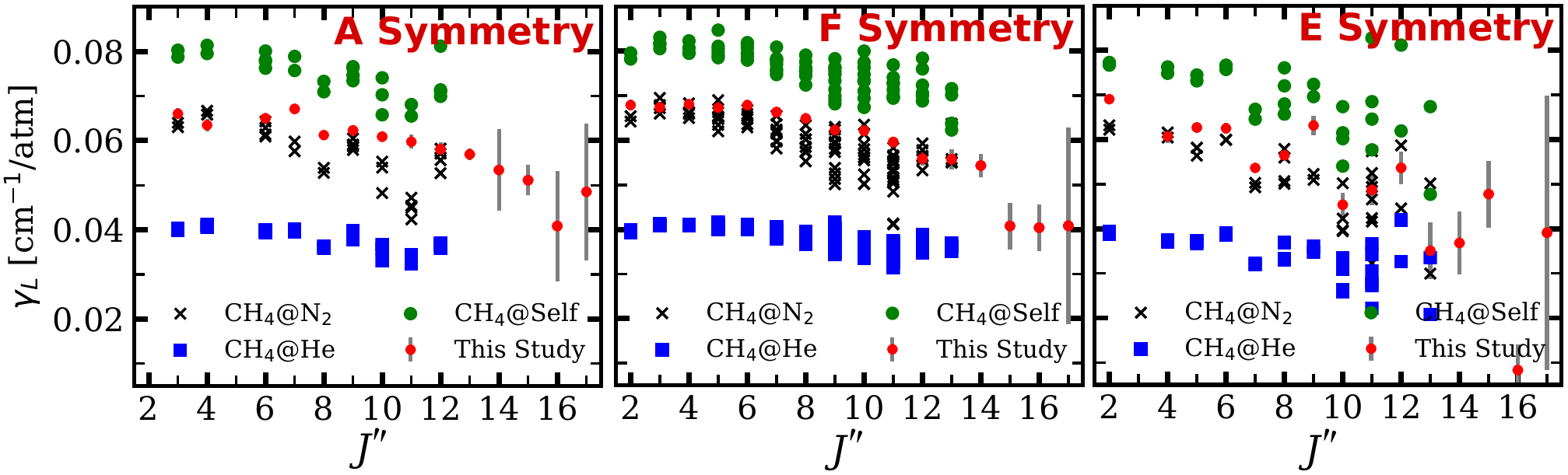}
 \caption{Comparison of our $\gamma_L$ results with the literature data including
   $\nu_3 \ Q$-branch \ce{CH4}@[\ce{He}, \ce{N2}, and Self] \citep{Pine1992}.
   Note that our results are averaged over ${\it N}$, however, the scatter of the literature data arises from the ${\it N}$ quantum index.
   { Where error bars are not visible, the uncertainties are smaller than the symbol size.}}
 \label{fig:compare-gamma}
\end{figure*}

Figure \ref{fig:compare-gamma} represents the effect of various broadeners (i.e., self or \ce{CH4}, \ce{N2}, and \ce{He}) on $\gamma_L$ for different $T_d$ symmetry species \cite{Pine2019_CH4_v3_theory_nonVoigt, Pine1992, Varanasi1989CH4nT,Varanasi1989ch4_v4, Es-sebbar2014_CH4_P_v3, Devi1983}.
In general  $\gamma_{\text{L}}$(Self) $>$ $\gamma_{\text{L}}$(\ce{H2}) $\geqslant$ $\gamma_{\text{L}}$(\ce{N2}) $>$ $\gamma_{\text{L}}$(\ce{He}).
{ In earlier work \cite{Tejwani1971}, electrostatic forces (dipole, quadrupoles, and higher-order multipoles) were theorised to cause the differing broadening effects of various broadeners (or perturbers) on \ce{CH4}. However, the quadrupole moments of \ce{O2} and \ce{N2} could not explain their similar broadening of \ce{CH4} (e.g., \cite{Devi1983}), given that their quadrupole moments differ by a factor of 3. Later, \citet{Neshyba1994}  showed that in fact atom-atom interactions supplant electrostatic interactions is a minor reason, and atom--atom interaction is the major source of broadening using Robert-Bonamy theory \cite{RobertBonamy1979} (see the theory section in Ref. \cite{Gabard2013}). }

\begin{figure*}[h] 
\centering
 \includegraphics[scale=1.0] {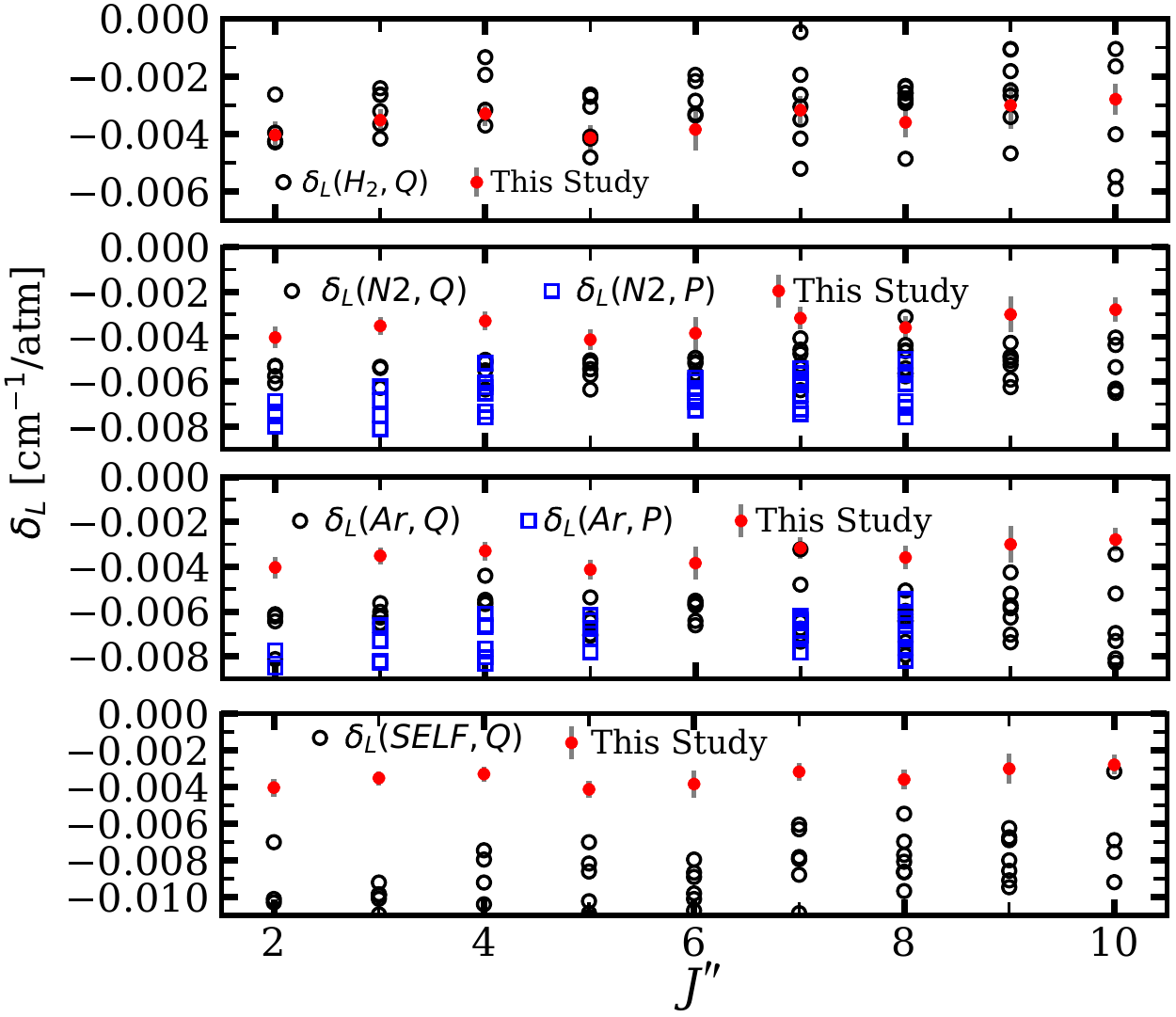} 
 \caption{The comparison of pressure-shift coefficient $\delta_{\text{L}}$ for different broadeners (i.e., \ce{H2}, \ce{N2}, Ar, and self-broadening), and different branches $Q$ and $P$ \cite{Pine1992, Pine1997, Pine2019_CH4_v3_theory_nonVoigt}. }
 \label{fig:compare-delta}
\end{figure*}

Figure \ref{fig:compare-delta} compares our Lorentzian pressure-shift results $\delta_{\text{L}} (\ce{CH4}@[\ce{H2}],P$ branch, see Table~\ref{tab:PB_ALL_J_Dependence}) with literature values for different broadeners and branches. The reported $\delta_{\text{L}}$ are averaged over of $J^{\prime\prime}$, and they are in the range of the \citet{Pine1992} $\nu_3$  $Q$-branch data. In addition, this figure shows that the collisional effect of \ce{N2} and \ce{Ar} species on pressure shift is larger than \ce{H2}. The largest $\delta_{\text{L}}$, however, would be due to the \ce{CH4} self-broadening interactions, and it is $\sim$2$\times$ higher than our results ($\delta_{\text{L}}(\ce{CH4}@[\ce{H2}])$).

\section{Summary $\&$ Conclusion}
High-temperature Lorentzian broadening and shift coefficients of \ce{CH4}@[\ce{H2}] for more than 100 individual rovibrational transitions in the $\nu_3$ $P$ branch are obtained using high resolution (0.01--0.005 cm$^{-1}$) FTIR spectroscopy. We find that $\gamma_{\text{L}}$ falls in the range 0.03--0.07 cm$^{-1}$/atm, and is strongly dependent on molecular rotation and symmetry dependent. The temperature-dependence broadening coefficient, $n_{\text{$\gamma$}}$ falls in the range 0.20--0.65. The averaged shift pressure and its temperature-dependence coefficient, $\delta_{\rm L}$ and $n_{\text{$\delta$}}$ are $-$0.0035 cm$^{-1}$/atm and  1.24, respectively, and these are constant with J as far as our data can determine.

All these coefficients were fitted to simple polynomial equations in terms of $J^{\prime\prime}$  and neglecting symmetry and $N$ quantum index  for the benefit of the astrophysical/exoplanetary community. Table S1 lists the $\gamma_{\rm {L}}$ and $n_{\text{$\gamma$}}$ for all individual lines, showing the change in these coefficients with $J$, symmetry, and ${\it N}$ numbers, and is recommended to use these data where these details are important. The detection of \ce{CH4} spectral features in hot-Jupiters to super-Earths needs these pressure-broadening data because of their  high-temperature and \ce{H2}-dominant atmospheres.       

These pressure-broadening and pressure-shift coefficients can be directly incorporated into current databases, such as HITRAN/HITEMP or EXOMOL. 
\vspace{-0.2cm}
\section{Acknowledgment}
\vspace{-0.2cm}
We kindly thank Glenn Stark, David Wright, Adam Schneider, and the Arizona State University exoplanet group for many useful discussions. E.G.N. especially thanks Mike Line for invaluable numerous invalvuable discussions during this work as well as Richard Freedman and Mark Marley invaluable discussions regarding the intricacies of opacity data. E.G.N. acknowledges funding from the GRSP research grant from the Arizona State University Graduate office program award XH51027. A.N.H.'s research was supported by an appointment to the NASA Postdoctoral Program at Arizona State University and the NASA Astrobiology Institute, administered by Universities Space Research Association under contract with NASA. This research used resources of the Advanced Light Source, which is a DOE Office of Science User Facility under contract no. DE-AC02-05CH11231.

\begin{table}[h]
  \caption{Broadening coefficients${\color{red}^\star}$ averaged over ${\it N}$.}
  \vspace{-0.3cm}
  \label {tab:PB_ALL_J_SYM_Dependence}
  \centering
  \footnotesize
\begin{tabular}{p{1pt}p{1pt}p{6pt}p{22pt}p{42pt}p{20pt}p{35pt}}

\hline
$J^{\prime}$  &  $J^{\prime \prime}$ & Sym & $\gamma_L$ [cm$^{-1}$/atm] & $\gamma_L^{min}$--$\gamma_L^{max}$  &  $n_{\text{$\gamma$}}$ & $n_{\text{$\gamma$}}^{min}$--$n_{\text{$\gamma$}}^{max}$ \\
\hline
1 &  2 &   $E$  &  0.069(1)    &      &   0.59(4)   &      \\ 
1 &  2 &   $F$  &  0.068(1)    &      &   0.66(3)   &       \\
2 &  3 &   $A$  &  0.066(1)    &      &   0.55(3)   &       \\
2 &  3 &   $F$  &  0.067(1)    &      &   0.57(2)   &       \\

3 &  4 &   $A$  &  0.064(2)    &      &   0.51(3)   &       \\
3 &  4 &   $E$  &  0.061(2)    &      &   0.56(4)   &      \\
3 &  4 &   $F$  &  0.068    &  0.068--0.069    &   0.60   &   0.60--0.62    \\

4 &  5 &   $E$  &  0.063    &  0.063--0.073    &   0.60   &  0.60--0.95    \\
4 &  5 &   $F$  &  0.067    &  0.065--0.072    &   0.60   &  0.54--0.68     \\

5 &  6 &   $A$  &  0.065    &  0.063--0.067    &   0.55   &   0.53--0.58    \\
5 &  6 &   $E$  &  0.063(1)    &      &   0.55(3)   &      \\
5 &  6 &   $F$  &  0.068    &  0.063--0.070    &   0.63   &  0.51--0.66     \\

6 &  7 &   $A$  &  0.067    &  0.067--0.069    &   0.65   &  0.59--0.65     \\
6 &  7 &   $E$  &  0.054(1)    &      &   0.51(2)   &      \\
6 &  7 &   $F$  &  0.066    &  0.064--0.068    &   0.60   &   0.57--0.62    \\

7 &  8 &   $A$  &  0.061(1)    &      &   0.56(2)   &       \\
7 &  8 &   $E$  &  0.057    &  0.056--0.059    &   0.54   &  0.52--0.53    \\
7 &  8 &   $F$  &  0.065    &  0.063--0.067    &   0.57   &  0.55--0.60     \\
  
8 &  9 &   $A$  &  0.062    &  0.061--0.079    &   0.53   &  0.52--0.88     \\
8 &  9 &   $E$  &  0.063    &  0.058--0.063    &   0.56   &  0.56--0.67    \\
8 &  9 &   $F$  &  0.062    &  0.061--0.064    &   0.56   &  0.51--0.60     \\
9 &  10 &  $A$  &  0.061    &  0.053--0.062    &   0.53   &  0.46--0.57     \\
9 &  10 &  $E$  &  0.045    &  0.045--0.062    &   0.44   &  0.07--0.45    \\
9 &  10 &  $F$  &  0.062    &  0.061--0.063    &   0.54   &  0.52--0.56     \\
10 & 11 &  $A$  &  0.060(2)    &      &   0.50(3)   &       \\
10 & 11 &  $E$  &  0.049    &  0.045--0.057    &   0.42   &  0.41--0.43    \\
10 & 11 &  $F$  &  0.060    &  0.055--0.059    &   0.50   &  0.32--0.59     \\
11 & 12 &  $A$  &  0.058    &  0.052--0.059    &   0.47   &  0.46--0.50     \\
11 & 12 &  $E$  &  0.054    &  0.049--0.054    &   0.41   &  0.40--0.62    \\
11 & 12 &  $F$  &  0.056    &  0.055--0.058    &   0.45   &  0.41--0.77     \\
12 & 13 &  $A$  &  0.057(1)    &      &   0.46(3)   &       \\
12 & 13 &  $E$  &  0.035    &  0.030--0.064    &   0.25  &  0.16--0.48     \\
12 & 13 &  $F$  &  0.056    &  0.055--0.062    &   0.48   &  0.36--0.60     \\
13 & 14 &  $A$  &  0.053(9)    &      &   0.54(18)  &        \\
13 & 14 &  $E$  &  0.037    &  0.031--0.048    &   0.19  &  0.19--0.29     \\
13 & 14 &  $F$  &  0.054    &  0.052--0.063    &   0.35  &  0.15--0.57     \\
14 & 15 &  $A$  &  0.051    &  0.049--0.055    &   0.35  &  0.22--0.50     \\
14 & 15 &  $E$  &  0.048    &  0.043--0.052    &   0.41  &  0.31--0.40     \\
14 & 15 &  $F$  &  0.041    &  0.027--0.059    &   0.14  &  0.15--0.53      \\
15 & 16 &  $A$  & 0.048${\color{red}^\dagger}$  &  0.041--0.068   &  0.24${\color{red}^\dagger}$  &   0.24--0.59     \\
15 & 16 &  $E$  &  0.027${\color{red}^\ddagger}$     &     &  0.24${\color{red}^\ddagger}$  &       \\
15 & 16 &  $F$  &  0.041     &  0.041--0.053   &   0.19  &    0.24--0.66    \\
16 & 17 &  $A$  &  0.049&  0.045--0.065   &   0.24  &   0.18--0.59     \\
16 & 17 &  $E$  &  0.039&  0.023--0.034   &   0.65  &   0.18--0.73    \\
16 & 17 &  $F$  &  0.038&  0.037--0.051   &   0.14  &   0.18--0.24     \\
\toprule
\toprule
\end{tabular}
 \scriptsize
 \parbox[t]{\linewidth}{Note:\\
${\color{red}^\star}$ The uncertainties in parentheses (in units of the least-significant digit) are derived from the estimated uncertainty of fitted linewidths.  These are not well defined where lines of differing quantum index, $N$, have been averaged and instead the range of parameters for individual lines is given as $\gamma_L^{min}$--$\gamma_L^{max}$ and $n_{\text{$\gamma$}}^{min}$--$n_{\text{$\gamma$}}^{max}$.
 ${\color{red}^\dagger}$Extracted values of  $\gamma_{\text{L}}$ and $n_{\text{$\gamma$}}$ for $J^{\prime \prime}$=16 are 0.041(12) and -0.00(27), which are out of the trend. Therefore, these values are replaced with the expected values from the polynomial equation~\ref{eq:fit-poly-J-sym} due to the weakness of the lines.\\
  ${\color{red}^\ddagger}$The extracted values of  $\gamma_{\text{L}}$ and $n_{\text{$\gamma$}}$ for $J^{\prime \prime}$=16 are 0.008(6) and -1.50(66), which are out of the trend. Therefore, these values are replaced with the expected values from the polynomial equation~\ref{eq:fit-poly-J-sym} due to the weakness of the lines.}
\end{table}

\clearpage
\newpage

\section{Reference}
\bibliographystyle{elsarticle-num-names}
\bibliography{reference}

\end{document}